\documentclass[fleqn,twoside]{article}

\usepackage{espcrc2}

\usepackage{graphicx}

\usepackage[figuresright]{rotating}

\newcommand{\AmS}{{\protect\the\textfont2

  A\kern-.1667em\lower.5ex\hbox{M}\kern-.125emS}}

\title{Exclusive Measurements of $pd\rightarrow ^3He\;\pi\pi$ :  the $ABC$
  Effect Revisited}

\author{M.~Bashkanov\address[PIT]{Physikalisches Institut der Universit\"at
  T\"ubingen, D-72076 T\"ubingen, Germany},
 D.~Bogoslawsky\address[JINR]{Joint Institute for Nuclear Research, Dubna,
  Russia},
H.~Cal\'en\address[SL]{The Svedberg Laboratory, Uppsala, Sweden},
F.~Cappellaro\address[UU]{Uppsala University, Uppsala,Sweden},
H.~Clement\addressmark[PIT],
L.~Demiroers\address[HU]{Hamburg University, Hamburg, Germany},
C.~Ekstr\"om\addressmark[SL],
K.~Fransson\addressmark[SL],
J.~Greiff\addressmark[SL],
L.~Gustafsson\addressmark[UU],
B.~H\"oistad\addressmark[UU],
G.~Ivanov\addressmark[JINR],
M.~Jacewicz\addressmark[UU],
E.~Jiganov\addressmark[JINR],
T.~Johansson\addressmark[UU],
O.~Khakimova\addressmark[PIT],
M.M.~Kaskulov\addressmark[PIT],
S.~Keleta\addressmark[UU],
I.~Koch\addressmark[UU],
F.~Kren\addressmark[PIT],
S.~Kullander\addressmark[UU],
A.~Kup\'s\'c\addressmark[SL],
A. Kuznetsov\addressmark[JINR],
P.~Marciniewski\addressmark[SL],
R.~Meier\addressmark[PIT],
B.~Morosov\addressmark[JINR],
W.~Oelert\address[FJ]{Forschungszentrum J\"ulich, Germany},
C.~Pauly\addressmark[HU],
Y.~Petukho\addressmark[JINR],
A.~Povtorejko\addressmark[JINR],
R.J.M.Y.~Ruber\addressmark[SL],
W.~Scobel\addressmark[HU],
T.~Skorodko\addressmark[PIT],
B.~Shwartz\address[BINP]{Budker Institute of Nuclear Physics, Novosibirsk,
  Russia},
V.~Sopov\address[ITEP]{Institute of Theoretical and Experimental Physics,
  Moscow, Russia},
J.~Stepaniak\address[SINS]{Soltan Institute of Nuclear Studies, Warsaw and
  Lodz, Poland},
V.~Tchernyshev\addressmark[ITEP],
P.~Th\"orngren-Engblom\addressmark[UU],
V.~Tikhomirov\addressmark[JINR],
A.~Turowiecki\address[IEP]{Institute of Experimental Physics, Warsaw, Poland},  G.J.~Wagner\addressmark[PIT],
M.~Wolke\addressmark[UU],
A.~Yamamoto\address[HEARO]{High Energy Accelerator Research Organization,
  Tsukuba, Japan},
 J.~Zabierowski\addressmark[SINS],
and
J.~Zlomanczuk\addressmark[UU]}

\begin{document}

\begin{abstract}

Exclusive measurements of the reactions $pd\rightarrow$ $^3$He $\pi^+ \pi^-$
and $pd\rightarrow$ $^3$He $\pi^0\pi^0$ have been carried out at $T_p=0.893$
GeV at the CELSIUS storage ring using the WASA detector. The $\pi^+\pi^-$
channel evidences a pronounced enhancement at low invariant $\pi\pi$ masses -
as anticipated from previous inclusive measurements of the ABC effect. This
enhancement is seen to be even much larger in the isoscalar $\pi^0\pi^0$
channel.
The differential distributions prove this enhancement to be of
scalar-isoscalar nature. $\Delta\Delta$ calculations give a good description of
the data, if a boundstate condition is imposed for the intermediate
$\Delta\Delta$ system. 

\vspace{1pc}

\end{abstract}


\maketitle

About 40 years ago first measurements onto the double-pionic fusion of protons
and deuterons to $^3$He particles led to a big surprise. In the momentum
spectra of the $^3$He particles detected by a magnetic spectrometer Abashian,
Booth and Crowe \cite{ABC} found an intriguing excess of strength close to the
$\pi\pi$ threshold. Follow-up measurements of this group suggested this
enhancement to be of isoscalar $I_{\pi\pi}=0$ nature, since corresponding
measurements on the
isovector $\pi^+\pi^0$ channel in $pd\rightarrow$ $^3$HX yielded a much smaller
cross section. Hence it has been speculated, whether some unknown isoscalar
resonance (like, e.g., the $\sigma$ meson) could be the origin of the observed
enhancement. Later on the effect, meanwhile referred to as ABC effect after
the initials of the original authors and interpreted as $\Delta\Delta$
excitation \cite{Anj}, was confirmed in much more detailed studies at Saclay
\cite{Ban}, even on other nuclear systems \cite{Wur,Col} - though in all cases
by inclusive single-arm magnetic spectrometer measurements detecting solely the
nuclear recoil particle.

The first exclusive measurements of the $pd\rightarrow$ $^3$He $\pi^+\pi^-$
reaction have been carried out recently at CELSIUS \cite{And} very close to
threshold and at COSY-MOMO \cite{Bel} near threshold. Whereas in the first
case the very limited statistics does not permit any definite conclusions, the
MOMO data clearly show a shift of strength towards high invariant $\pi^+\pi^-$
masses $M_{\pi^+\pi^-}$ as compared to phase space distributions. Though this
finding near threshold is in contrast to the ABC effect observed at much
higher energies, it coincides with that in $pp\rightarrow pp\pi^+\pi^-$
\cite{Bro}, where the close-to-threshold data are in accordance with
excitation and decay of the Roper resonance.

In order to shed more light on the nature of the ABC effect we have carried
out exclusive measurements of the reactions $pd\rightarrow$ $^3$He $\pi^0$,
$^3$He $\pi^0\pi^0$ and $^3$He $\pi^+\pi^-$ at $T_p=0.893$ GeV using the WASA
detector  
\cite{Zab} with the deuterium pellet target system at the CELSIUS storage
ring. The energy chosen corresponds to the one, where the maximum ABC effect
has been observed \cite{Ban}. The detector has nearly full angular coverage
for the detection of charged and uncharged particles. The forward detector
consists of a thin window plastic scintillator hodoscope at the exit of the
scattering chamber, followed by straw tracker, plastic scintillator quirl and
range hodoscopes, whereas the central detector comprises in its inner part a
thin-walled superconducting magnet containing a minidrift chamber for tracking
and in its outer part a plastic scintillator barrel surrounded by an
electromagnetic calorimeter consisting of 1012 CsI (Na) crystals. 

\begin{figure}
\begin{center}
\includegraphics[width=18pc]{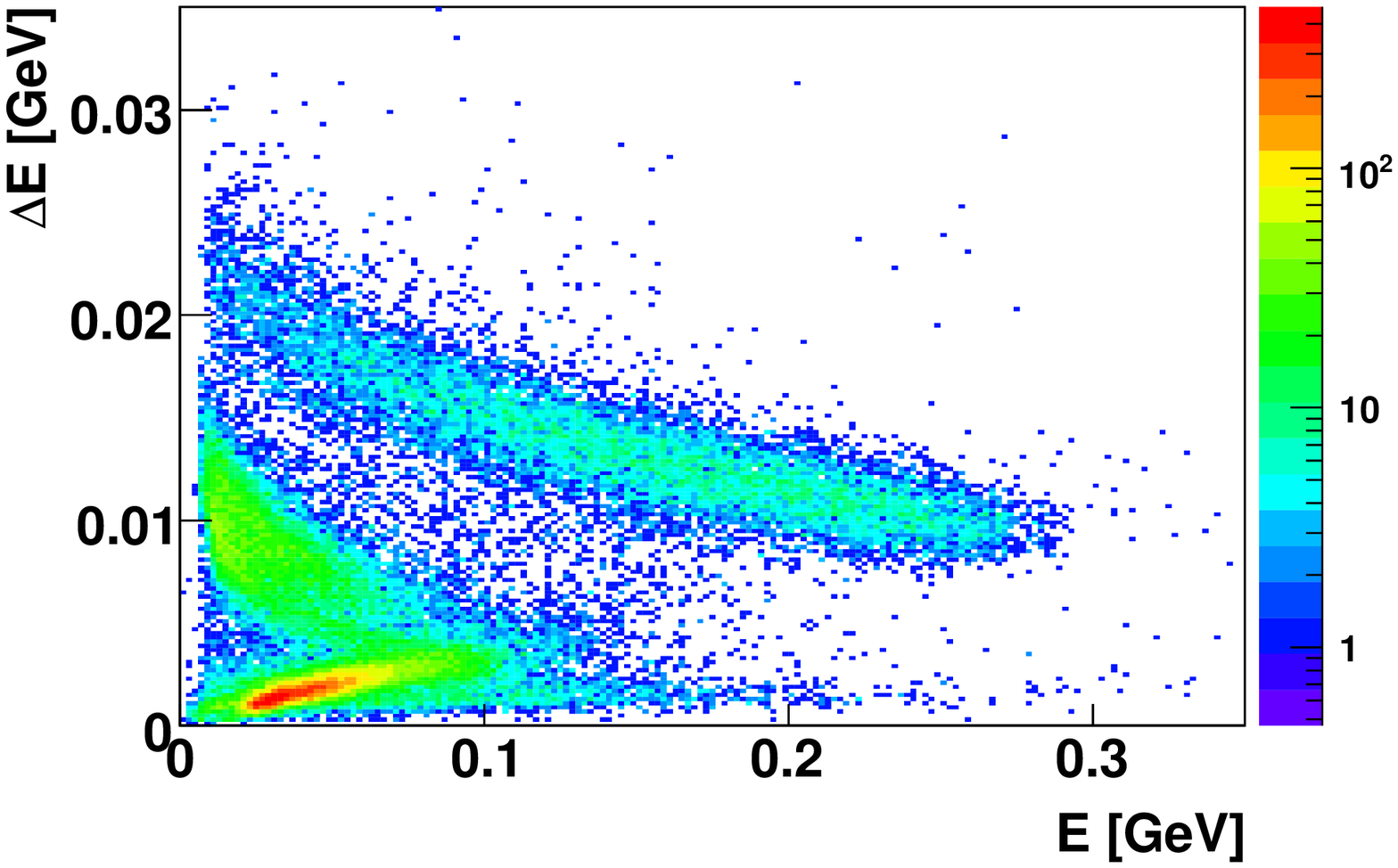}
\includegraphics[width=18pc]{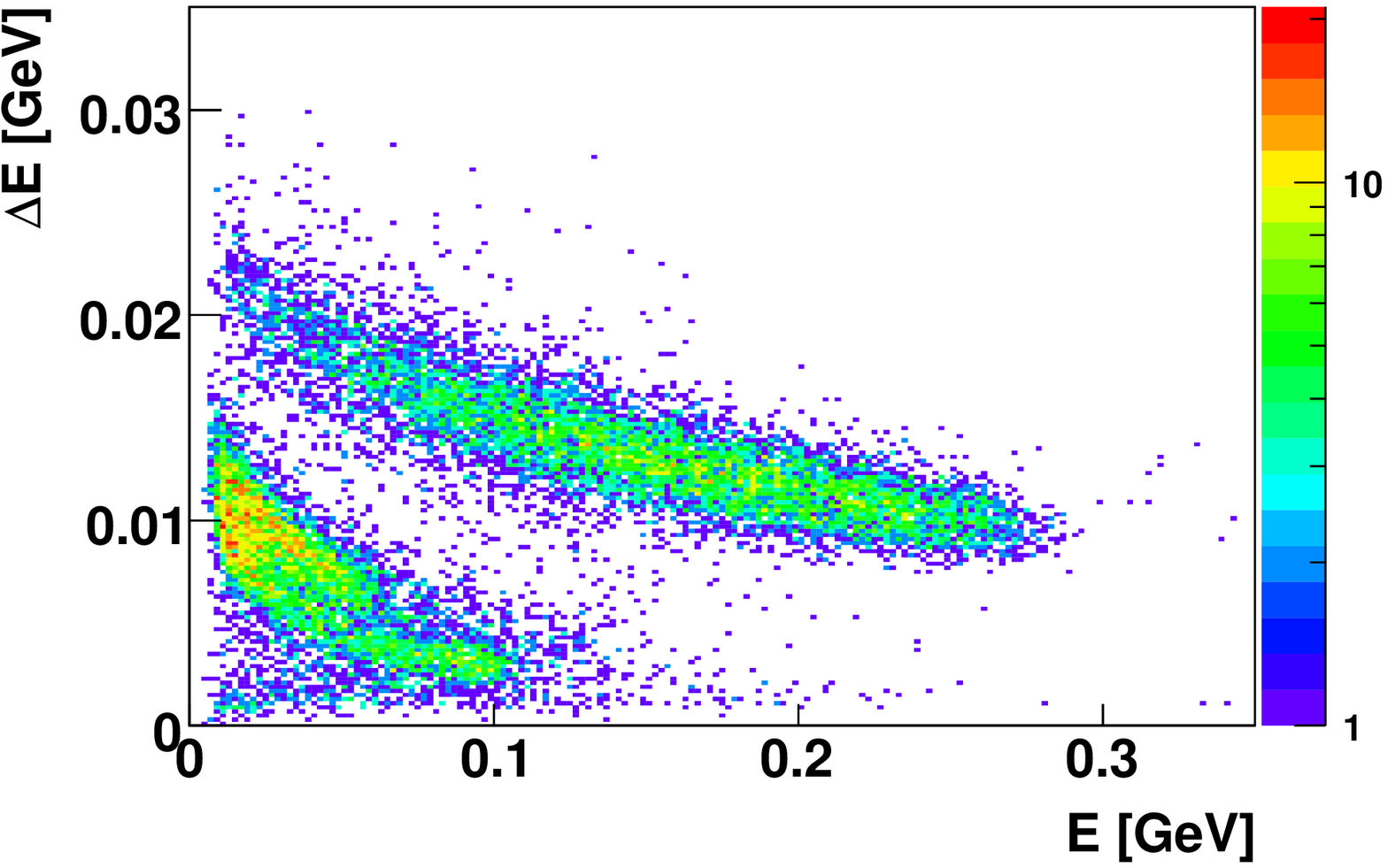}
\end{center}
\caption{$\Delta$E-E scatterplot for particles stopped in the forward
  detector, on top without any constraints, at
  bottom with the constraint of an identified $\pi^+\pi^-$ or $\pi^0\pi^0$
  pair in the central detector. Here $\Delta$E and E denote the energy loss in
  the third layer
  of quirl and range hodoscope, respectively. The upper band in the plots
  shows the recorded $^3$He particles, whereas the lower band contains mainly
  protons and deuterons. Note also that due to the constraint
  $\Theta^{cm}_{^3He} \leq 90^\circ$ only $^3$He particles with $E \ge 0.1$ GeV
  are taken for the further analysis. } 
\end{figure}

$^3$He
particles have been detected in the forward detector and identified by the
$\Delta$E-E technique using corresponding informations from quirl and
range hodoscope, respectively. In order to suppress the vast background of
fast protons and other minimum ionizing particles already on the
trigger level, appropriate $\Delta$E thresholds have been set on the window
hodoscope acting as a first level trigger.
Fig. 1, top, shows the $\Delta$E-E scatterplot of the events registered in the
forward detector. The $^3$He band is clearly separated from that for deuterons
and protons. 

Charged pions and gammas (from $\pi^0$ decay) have been detected in the
central detector. This way the full four-momenta have been measured for all
particles of an event allowing thus kinematic fits with 4 overconstraints in
case of $\pi^+\pi^-$ production and 6 overconstraints in case of $\pi^0\pi^0$
production. Fig.1, bottom, shows the  $\Delta$E-E scatterplot for events,
which contain a pion pair identified in the central detector. There is
practically no background any longer beneath the $^3$He band.

\begin{figure}
\begin{center}
\includegraphics[width=16pc]{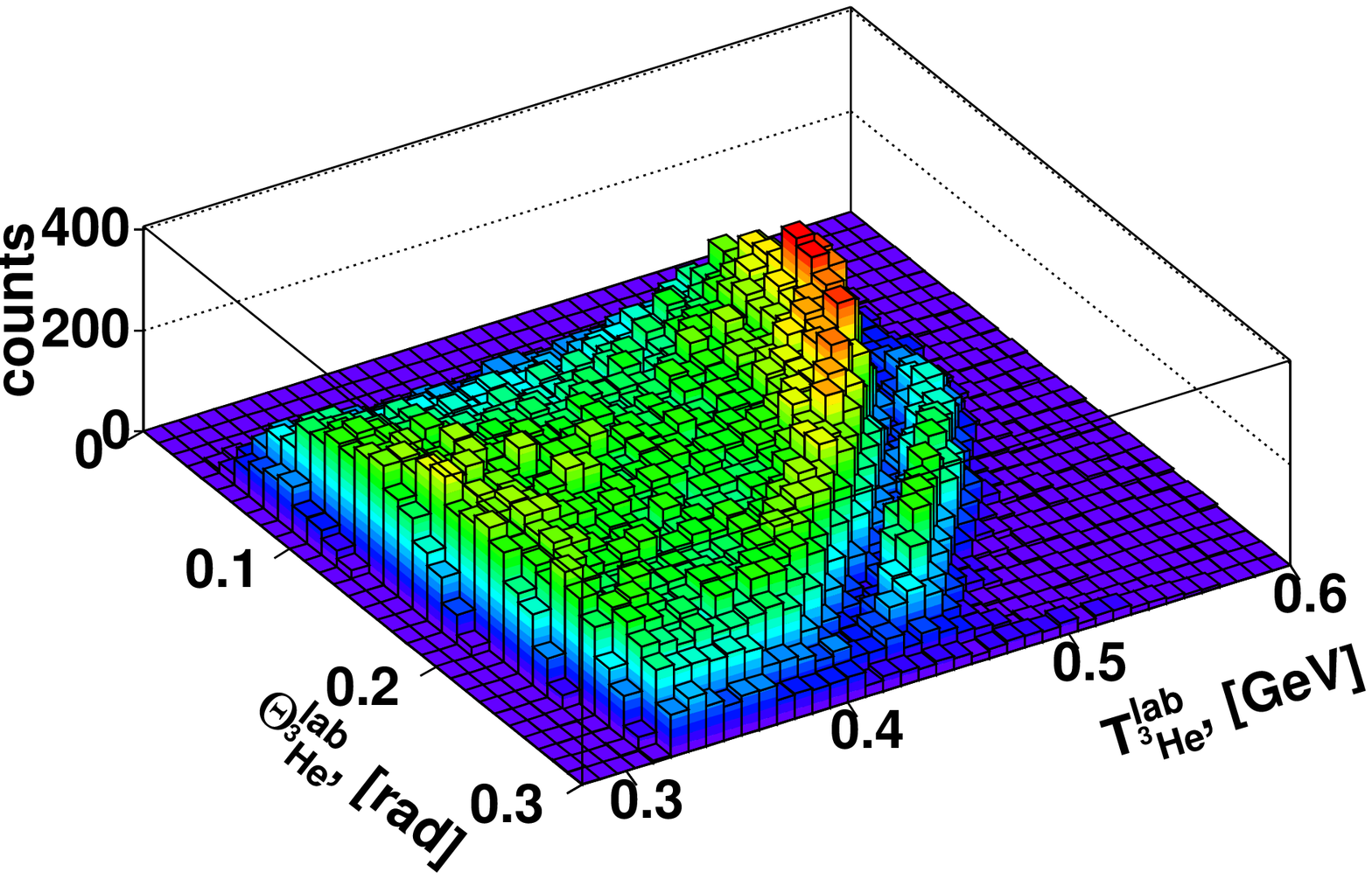}
\includegraphics[width=16pc]{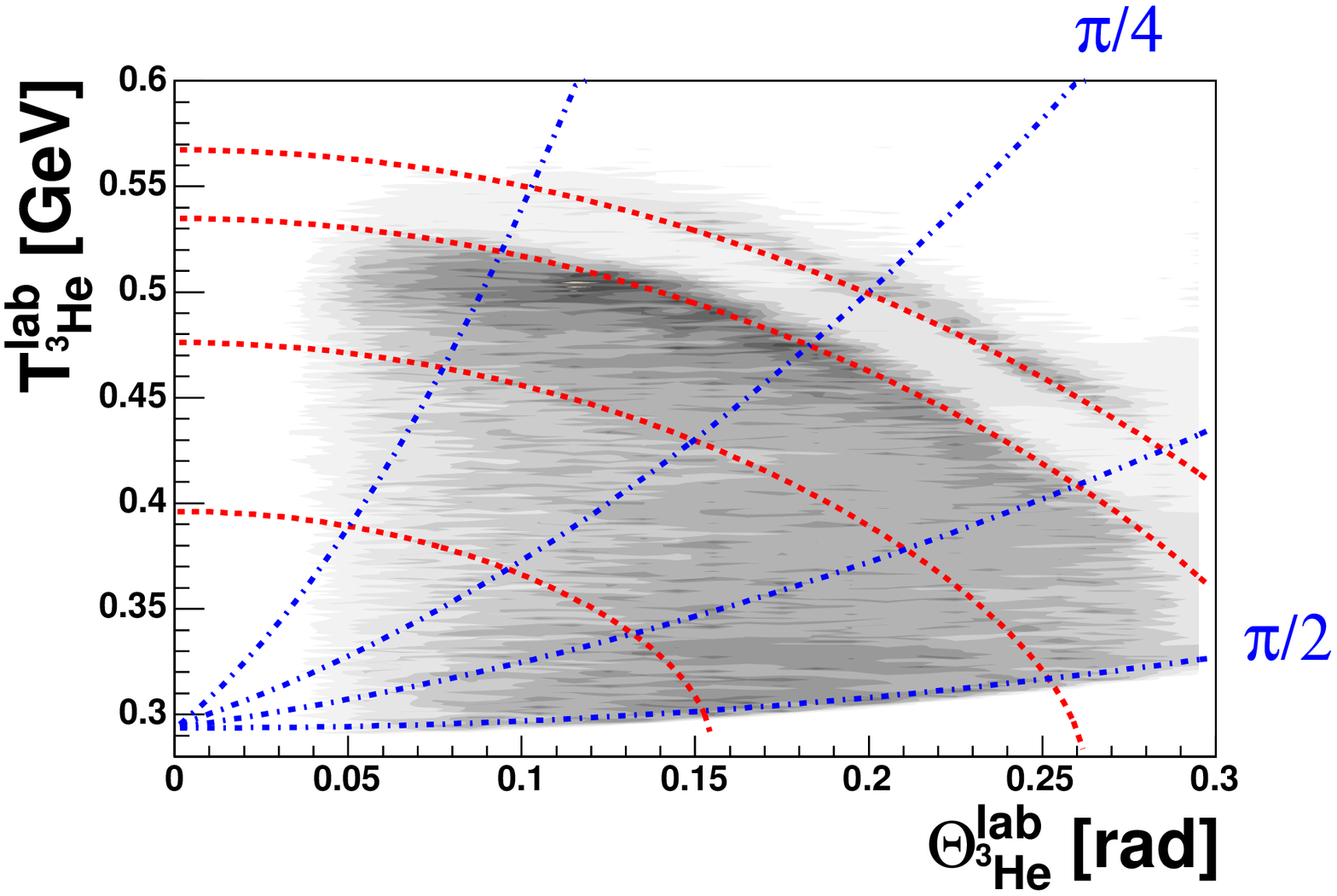}
\end{center}
\caption{3D and contourplots of lab angle $\Theta^{lab}_{^3He}$ versus lab
  energy $T^{lab}_{^3He}$for $^3He$ particles measured in the forward
  detector. The dash-dotted lines give $\Theta^{cm}_{^3He}=22.5^\circ,
  45^\circ,  67.5^\circ, 90^\circ$.  The dashed lines indicate the contours of
  missing masses $MM_{^3He} =$ 0.135, 0.27, 0.4, 0.5 GeV} 
\end{figure}

In order to faciltate comparison with the previous inclusive measurements
\cite{Ban} we
display in Fig. 2 lego and contour plots of lab angle versus lab energy of
the $^3$He particles detected in the forward detector - before kinematic fit
and any demand on other particles in the event. Whereas for single $\pi^0$
production $^3$He particles have been registered only in a very limited angle
and energy range of phase space, the $^3$He particles stemming from $\pi\pi$
production have been detected over the full kinematical range
up to $^3$He angles $\Theta^{cm}_{^3He} \leq 90^\circ$. Since we demand the
$^3$He particles to reach the range hodoscope they need to have kinetic
energies of more than 200 MeV in order to be registered and savely
identified. Hence for $^3$He cms angles much larger than $90^\circ$ the phase
space is no longer fully covered in our measurement. In order to avoid
extrapolations we introduce therefore the $90^\circ$ cut.

In Fig. 2 the band for
single $\pi^0$ production is seen to be well separated from the continuum for
$\pi\pi$ production. Also immediately evident is a large accumulation of
events near the kinematical limit for $\pi\pi$ production, i.e. in the region
corresponding to small invariant $\pi\pi$ masses. Since the detector efficiency
is approximately constant over the corresponding phasespace region in Fig.2,
this feature obviously is in accord with a strong ABC enhancement present in
these data. In fact, if we divide Fig. 2 into angular bins, then we obtain
spectra in resemblance of those measured at Saclay \cite{Ban}. As an
example we show
in Fig. 3 the $^3$He momentum spectrum for the angular bin  $7^\circ \leq
\Theta^{lab}_{^3He} \leq 8^\circ$ . The rise at small momenta - corresponding
to large missing masses - in the inclusive spectrum can be associated with
$\pi\pi\pi$ production (dash-dotted histogram in Fig.3) production and 
$I_{\pi\pi}=1$ contributions as will become evident from the analysis of the
exclusively measured data.

\begin{figure}[t]
\begin{center}
\includegraphics[width=16pc]{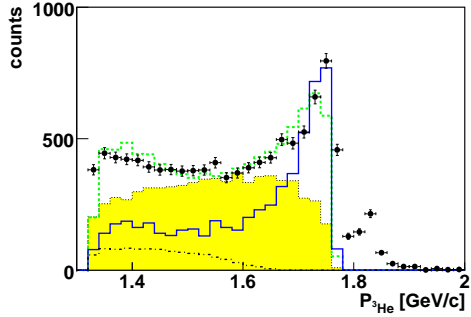}
\end{center}
\caption{$^3$He momentum spectrum for the angular bin  $7^\circ \leq
\Theta^{lab}_{^3He} \leq 8^\circ$ (not corrected for detector efficiency). The
data points represent the
inclusively measured spectrum for comparison with the Saclay data
\cite{Ban}. The shaded histogram displays the phase space for $\pi\pi$
production normalized as to touch the data as in Ref.\cite{Ban}. The
dash-dotted histogram shows the phase space calculation for $\pi\pi\pi$
production. The dashed histogram represents conventional $\Delta\Delta$
calculations normalized to the data, whereas the solid histogram shows the same
calculations with a boundstate condition for the $\Delta\Delta$ system
included.
} 
\end{figure}

Next we demand that the $^3$He particles are accompanied with 2 or 4 gammas
from $\pi^0$ decay or a $\pi^+\pi^-$ pair registered in the central
detector. The completeness of the events is checked by appropriate missing
mass conditions. Having now also the four-momenta of the accomponying particles
the selected events are overcomplete and kinematical fitting with 4 to 6
overconstraints can be applied. To obtain further information on the $\pi\pi$ 
production process we next consider differential observables after correction 
of the data for acceptance and efficiency. The absolute normalisation of the
$\pi\pi$ data is done by normalizing our $^3$He $\pi^0$ data to those from
Saclay measurements \cite{Ban1} at neighboring energies. This gives a total
cross section of 2.8 (3) $\mu b$ for the $\pi^0\pi^0$ production and 5.1 (5)
$\mu b$ for the $\pi^+\pi^-$ production. The quoted uncertainties being
more than twice the purely statistical ones constitute a
very conservative estimate of systematic uncertainties particularly in the
absolute efficiency of the detector based on Monte Carlo
simulations of the detector performance.

\begin{figure}
\begin{center}
\includegraphics[width=16pc]{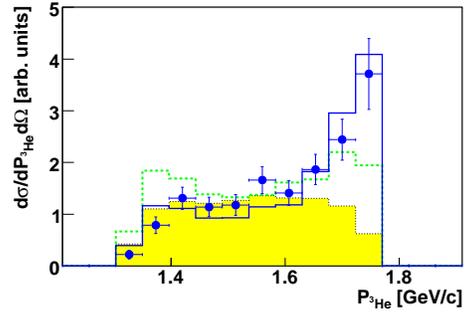}
\end{center}
\caption{The same as Fig.3, but now for the exclusively measured and
  efficiency corrected $\pi^0\pi^0$ channel.
} 
\end{figure}

 Note that the $\pi\pi$ 
production cross section is considerably larger than that for
single $\pi^0$ production (2.7 $\mu b$, averaged value from Ref.\cite{Ban1}
) as already apparent from Fig. 2. In the free NN system at corresponding
energies $\pi^0$ production is larger than $\pi^0\pi^0$ production by
more than an order of magnitude. The strong suppression of single $\pi^0$
production here is due to the constraint of having all nucleons in the exit
channel fused to $^3$He. 

As mentioned before we cover the full kinematical range of $\pi\pi$ production
only for $\Theta^{cm}_{^3He} \leq 90^o$. Hence we plot observables in the
following only for this kinematical region. If the $^3$He angular
distributions indeed are symmetric about 90$^\circ$ as indicated in the Saclay
measurements \cite{Ban}, then our data in fact do contain the complete
information over the full phase space. The numbers quoted above for the total
cross sections actually are derived under this assumption. 


Fig. 4 shows the $^3$He momentum distribution for the same angular slice as in
Fig. 3, however, selecting now only the exclusively measured $\pi^0\pi^0$
channel. We see that the relative enhancement at
high momenta has strongly increased, whereas the rise at low momenta has
disappeared. We note that already in Ref. \cite{Col} it had been suggested
that the low-momentum rise in the inclusive spectra might be due to $\pi\pi\pi$
rather than $\pi\pi$ production.

Figs. 5 and 6 display the differential distributions for
the invariant masses $M_{\pi\pi}$ and $M_{^3He \;\pi}$  as well as angular
distributions for $\delta_{\pi\pi}$, the opening angle between the two pions,
for the angle of the total momentum of $\pi\pi$ system
$\Theta^{cm}_{\pi\pi}=-\Theta^{cm}_{^3He}$ - all in the overall cms - and 
for  $\Theta^{\pi\pi}_{\pi}$, the pion angular distribution in the
$\pi\pi$ subsystem (Jackson frame), which is shown both for the full mass
region (circles) and for $M_{\pi\pi} < $ 0.34 GeV (squares).  
The shaded areas in these plots give the pure phase space distributions. Note
that for an (unpolarized) three-body reaction there are only 4 independent
observables.

As expected from the discussion of Figs. 1 - 4 we see a strong enhancement
in $M_{\pi^0\pi^0}$ at low masses towards threshold. In $M_{\pi^+\pi^-}$ we
see such an enhancement, too, however of considerably smaller size. Since low
invariant masses belong to $\pi\pi$ pairs with small relative momentum, such
pairs must move essentially in parallel both in lab and overall cms thus
having a small opening angle $\delta_{\pi\pi} \to 0$ as indeed is borne out in
our data for  $\delta_{\pi\pi}$ (Fig. 6). The $M_{^3He \;\pi}$ spectra exhibit
clear signatures of $\Delta$ excitation in the course of the reaction process
as anticipated by $\Delta\Delta$ calculations.

The $\Theta^{cm}_{^3He}$ angular dependence is similar to the Saclay results
( Figs. 20 and 21 in \cite{Ban}) though in detail somewhat flatter. However,
at Saclay only the 
angular dependence of the visible enhancement (ABC peak) in the momentum
spectra was looked at. 

For $\Theta^{\pi\pi}_\pi$ we obtain a slightly non-flat angular distribution
in the $\pi^0\pi^0$ channel. 
For $M_{\pi\pi} <$ 0.34 GeV, i.e. the region of the enhancement, the
$\Theta^{\pi\pi}_\pi$ distribution turns out, however, to be flat in
accordance with pure s-waves, which means that the enhancement is of scalar
nature. Note that, since the two $\pi^0$ cannot be distinguished, the angular
distribution has to be symmetric about $90^\circ$ by construction. In order to
treat the $\pi^+\pi^-$ channel on the same footing we also sort the charged
pions independently of their charge into the $\Theta^{\pi^+\pi^-}_\pi$ spectrum
symmetrizing thus this spectrum, too.

\begin{figure}[t]
\begin{center}
\includegraphics[width=18pc]{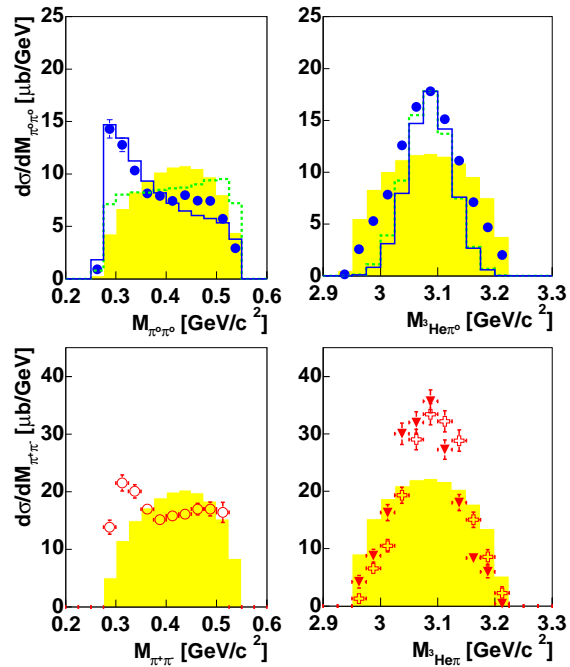}
\end{center}
\caption{Differential cross sections for the distributions of the invariant
  masses $M_{\pi\pi}$ (left) and $M_{^3He \;\pi}$ (right, with filled triangles
  for $M_{^3He \;\pi^-}$ and open crosses for $M_{^3He \;\pi^+}$) . Top:
  $pd\rightarrow$ $^3$He 
  $\pi^0\pi^0$, bottom: $pd\rightarrow$ $^3$He $\pi^+\pi^-$.
The shaded areas show the phase space distributions for comparison. Solid and
 dashed curves denote $\Delta\Delta$ calculations with and without boundstate
 condition.}
\end{figure}

\begin{figure}[t]
\begin{center}
\includegraphics[width=18pc]{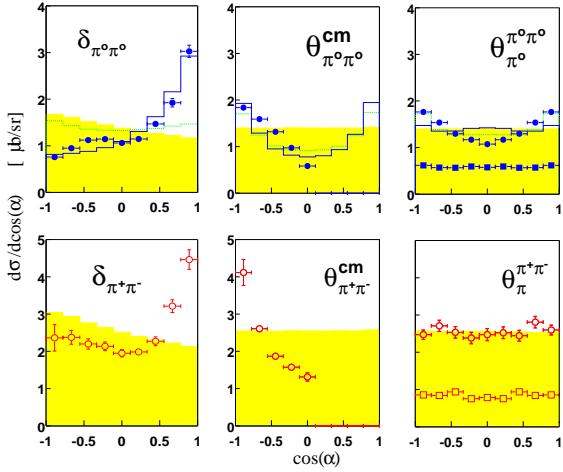}
\end{center}
\caption{Angular distributions of  
 the opening angle $\delta_{\pi\pi}$ between the two pions, the angle of the
 total momentum of the 
 $\pi\pi$ system $\Theta^{cm}_{\pi\pi}=-\Theta^{cm}_{^3He}$ - all in
 the overall cms - as well as the pion angle $\Theta^{\pi\pi}_{\pi}$ in the
 $\pi\pi$ subsystem  (Jackson frame). For the latter the data are plotted also
 with the  constraint $M_{\pi\pi} < 0.34$ GeV. Top:
  $pd\rightarrow$ $^3$He 
  $\pi^0\pi^0$, bottom: $pd\rightarrow$ $^3$He $\pi^+\pi^-$. For the meaning of
 symbols and curves see caption of Fig. 5.}
\end{figure}


Since the initial deuteron has isospin 0, whereas proton and $^3$He have
isospin 1/2, the reaction acts as an isospin filter and the $\pi\pi$ pair can
only be in an $I_{\pi\pi}=0$ or 1 state - in case of $\pi^0\pi^0$ even
uniquely in $I_{\pi\pi}=0$ 
due to Bose symmetry. From previous $pd\rightarrow$ $^3$H $\pi^+\pi^0$
measurements \cite{ABC,Ban}, where $I_{\pi\pi}=1$ solely, the
isovector contribution in $pd\rightarrow$ $^3$He $\pi^+\pi^-$ has been deduced
to be smaller by roughly an order of magnitude using isospin relations between
both channels. In addition Bose symmetry necessitates the two pions in the 
isovector state to be in relative p-wave, which in turn is
suppressed at small invariant masses. Therefore the ABC enhancement has
been assigned to be of isoscalar nature. On the other hand 
an $I_{\pi\pi}=1$ contribution in the $\pi^+\pi^-$
channel should show up at large invariant masses. To investigate this point in
more detail we plot both distributions on top of each other in Fig. 7 and
divide the  $\pi^+\pi^-$ cross section by a factor of two in order to account
for the isospin factor between both channels in case of $I_{\pi\pi}=0$. (
Actually this factor needs still to be corrected for the different reaction
thresholds of both channels due to the different masses of charged and neutral
pions. This kinematical correction is substantial at incident energies near
the reaction threshold \cite{And}, however close to unity at our energy far
above threshold, where the cross section saturates.) We
then see that  $\sigma (\pi^+\pi^-) / 2 \approx \sigma (\pi^0\pi^0)$ in the
region 320 MeV $\leq  M_{\pi\pi} \leq$ 400 MeV and somewhat larger beyond. If
we assign this surplus to the  $I_{\pi\pi}=1$ contribution then we end up with
0.6 $\mu b$ for the isovector part. The exact value, however, depends strongly
on the relative normalizations of $\pi^+\pi^-$ and $\pi^0\pi^0$
channels. Accounting for their uncertainties given above we could have an
isovector contribution as large as 1 $\mu b$ in support of the
findings in the inclusive measurements \cite{ABC,Ban}.

Coming back to the low-mass region we see (Fig. 7) that strength
starts to build up towards low masses in $M_{\pi^+\pi^-}$ until it gets
cut off by the $\pi^+\pi^-$
threshold. Due to the $9$ MeV lower $\pi^0\pi^0$ threshold,
the strength can continue growing towards smaller masses in the $\pi^0\pi^0$
channel. So the main discrepancy between both channels in their low-mass
behavior may possibly be associated with the
different thresholds, which in turn arise from the different $\pi^\pm$ and
$\pi^0$ masses and the isospin breaking residing in them. Coulomb effects
within the $\pi^+\pi^-$ system can be estimated by the
Gamov factor \cite{Jur} giving a 3$\%$ effect at  $M_{\pi^+\pi^-} = 0.3$ GeV,
which is within the uncertainties of the data and small compared to
the discrepancies between $M_{\pi^+\pi^-}$ and $M_{\pi^0\pi^0}$ spectra.
Coulomb effects between the  $\pi^+\pi^-$ pair and the $^3$He nucleus could
be somewhat larger due to the doubly charged nucleus, but have not been
calculated to our knowledge. 


\begin{figure}
\begin{center}
\includegraphics[width=18pc]{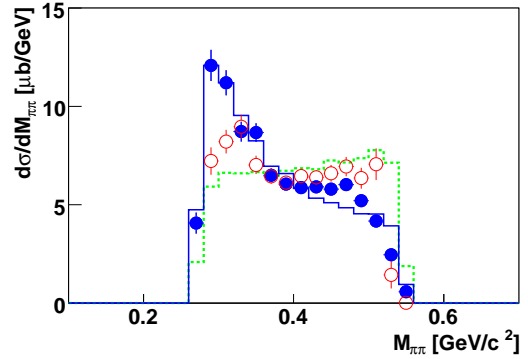}
\end{center}
\caption { Comparison of the $M_{\pi\pi}$ distribution for $\pi^0\pi^0$
  (full symbols) and $\pi^+\pi^-$ (open symbols) channels. The chosen binning
  reflects the experimental resolution. For easy comparison the 
  $\pi^+\pi^-$ cross section has been divided by a factor of two - the isospin
  factor relating both channels in the isoscalar case. For the meaning of the
  curves see caption of Fig. 5.
}
\end{figure}




We have not yet discussed possible reasons for the observed low-mass
enhancement as it manifests itself particularly well in the $M_{\pi^0\pi^0}$
spectrum as the clean isoscalar case. $\Delta\Delta$ calculations
\cite{Anj} for the isoscalar channel, which were 
thought to provide an explanation for the ABC effect on basis of inclusively
measured spectra, give a modest double-hump structure ( relative to phase
space ) in $M_{\pi\pi}$, which obviously is not supported any longer by the
new exclusively measured data. A double-hump structure appears to
be an important feature of $\Delta\Delta$ excitations, since to our knowledge
all calculations to this issue for bound  \cite{Gar,Alv} nuclear systems in
the final state  predict such a behavior. Indeed, it has been demonstrated in
the very basic and pedagogical paper of Risser and Shuster \cite{ris} that the 
$\Delta\Delta$ amplitude produces this structure by favoring the
configurations with the two pions moving
either in parallel (low-mass peak) or in antiparallel (high-mass peak), while
the two nucleons get confined in
the final nuclear bound state. In order to demonstrate this situation we have
carried out $\Delta\Delta$ calculations along the lines of Ref.\cite{ris}
including in addition appropriate angular dependences for the N$\pi$ system
in the $\Delta$ state. These calculations are shown in Figs. 3 - 7
by the dashed histograms. They, indeed, describe amazingly well the
inclusively measured $^3$He momentum spectra, however, are at variance with
the exclusive data for momentum and $M_{\pi\pi}$ spectra. Note that these
calculations, though creating big high- and low-momentum enhancements in the
momentum spectra, lead to enhancements in $M_{\pi\pi}$, which look quite
modest there due to the parabolic shape of the phase space distribution in
this spectrum. It has been shown \cite{Alv} that this double-hump structure
gets strongly enhanced in calculations, which account for $\rho$ exchange and
short range correlations in addition to $\pi$ exchange. 

Since obviously calculations, which predict a pronounced low-mass enhancement
in $M_{\pi\pi}$, also predict an even larger high-mass enhancement, which
however is
absent in the data for the isoscalar channel, an important piece appears to be
missing in such calculations. In the basic picture presented in Ref.\cite{ris}
this means that the configuration with the two pions moving in antiparallel
obviously is suppressed by some reason, e.g., if the two
$\Delta$ are hindered in their relative motion. In this case, since the two
nucleons of the two $\Delta$ are already confined by the nuclear boundstate
condition, also the two pions are forced to minimum relative
motion. Qualitatively this has the same effect in $M_{\pi\pi}$ as a strong
$\pi\pi$ final state interaction originally introduced for the explanation of
the ABC effect \cite{ABC}. Indeed,
if we impose such a condition for a quasi-bound $\Delta\Delta$ system in the
$\Delta\Delta$ calculations by a simple Gaussian formfactor, then we can
obtain a reasonable description of all data for the isoscalar $\pi^0\pi^0$
channel (Figs. 4 - 7, solid histograms). We note that there are quark model
based calculations, which indeed predict bound $\Delta\Delta$ states
\cite{fan}.  

Another possibility might be chiral
restoration \cite{Hat} \cite{Aou} or other dynamic  \cite{Roc}  effects in the
nuclear medium as observed, e.g., in $\pi A\rightarrow \pi\pi X$ \cite{Bon}
\cite{Sta} and $\gamma A\rightarrow \pi\pi X$  \cite{Mes} reactions, or a
resonance-like phenomenon near $\pi\pi$ 
threshold, which - since not observed in reactions on a single nucleon - would
need to be associated with the NN system. Such a phenomenon has been noted,
e.g., in Refs. \cite{San,Ani}. 

Summarizing, the first exclusive data on $pd\rightarrow$$^3$He $\pi^0\pi^0$
and $pd\rightarrow$$^3$He $\pi^+\pi^-$ in the ABC region reveal a strong
enhancement in the low $M_{\pi\pi}$ region - in the $\pi^0\pi^0$ channel being
much larger than in the $\pi^+\pi^-$ channel. From the data we see that this
enhancement is of scalar-isoscalar nature and hence shows up especially
pronounced in the $\pi^0\pi^0$ channel. 
The reason for the
low-mass enhancement appears still to be an unsettled problem. Ideas presented
include medium effects or some unknown phenomenon associated with the
NN-system - or possibly a $\Delta\Delta$ mechanism substantially different
from what has been assumed previously and possibly connected with a
quasi-bound  $\Delta\Delta$ system.

We are grateful to the TSL/ISV personnel for the continued help during the 
course of these measurements as well as to Colin Wilkin, Eulogio Oset and
Fan Wang for
valuable discussions on this matter. This work has been supported by BMBF
(06TU201), 
DFG (Europ. Graduiertenkolleg 683), Landesforschungsschwerpunkt
Baden-W\"urttemberg and the Swedish Research Council. We also acknowledge the
support from 
the European Community-Research Infrastructure Activity under FP6
"Structuring the European Research Area" programme (Hadron Physics, contract
number RII3-CT-2004-506078).

\end{document}